\documentclass[twoside]{ilcws08}
\usepackage[latin1]{inputenc}
\usepackage[dvips]{graphicx,epsfig,color}
\usepackage{wrapfig,rotating}
\usepackage{amssymb,amsmath,array}

\begin{document}
\title{
NLO Electroweak Corrections to Higgs Decay to Two Photons} 
\author{Stefano Actis
\thanks{ Work supported by the DFG through SFB$/$TR9 and by the EC MCRTN
         HEPTOOLS. Report numbers: PITHA 09$/$02; SFB$/$CPP-09-06. Slides
         available at~\cite{url}. }
\vspace{.3cm}\\
Institut f\"ur Theoretische Physik E, RWTH Aachen University, \\
D-52056 Aachen - Germany
}

\maketitle

\begin{abstract}
  The recent calculation of the next-to-leading order electroweak 
  corrections to the decay of the Standard Model Higgs boson 
  to two photons in the framework of the complex-mass 
  scheme is briefly summarized.
\end{abstract}

\section{Introduction}

The production of the Standard Model Higgs boson in photon-photon 
collisions represents an interesting mechanism for measuring the 
partial width $\Gamma(H\to\gamma\gamma)$ with a $3 \%$ accuracy at 
an upgrade option of the International Linear Collider (see~\cite{Monig:2007py}
and references therein).
Therefore, the computation of radiative corrections to the leading order 
(LO) decay width~\cite{Ellis:1975ap} has been an active field of research 
in the last years.

QCD corrections to the partial width of an intermediate-mass Higgs
boson have been computed at next-to-leading order (NLO) in~\cite{Zheng:1990qa} 
and at next-to-next-to-leading order (NNLO) in~\cite{Steinhauser:1996wy}.
The NLO result has been extended in~\cite{Melnikov:1993tj} to the entire 
Higgs-mass range.

Electroweak NLO corrections have been evaluated in~\cite{Liao:1996td} 
in the large top-mass scenario and in~\cite{Korner:1995xd} assuming a 
large Higgs-mass scenario. The two-loop corrections due to light fermions 
have been derived by the authors of~\cite{Aglietti:2004nj}, and electroweak 
effects due to gauge bosons and the top quark have been evaluated through an
expansion in the Higgs external momentum in~\cite{Degrassi:2005mc}.

More recently, all electroweak corrections at NLO have been computed
in~\cite{Passarino:2007fp} for a wide range of the Higgs mass, including 
the region across the $W$-pair production threshold. It has been shown
that the divergent behavior of the amplitude at a two-particle normal
threshold can be removed introducing the complex-mass scheme
of~\cite{Denner:2005es}, a program carried out in~\cite{Actis:2008ts}.
Note that we have not addressed the issue of re-summing Coulomb
singularities, as performed by the authors of~\cite{Melnikov:1994jb}.

In Section~\ref{sec:sings} of this note we review the implementation
of the complex-mass scheme in the two-loop computation and the effect
on the two-particle threshold region. Next, in Section~\ref{sec:num}, 
we discuss the numerical impact of the NLO electroweak corrections
on the partial width $\Gamma(H\to\gamma\gamma)$.

\section{Normal thresholds and complex masses}
\label{sec:sings}

The $H\to\gamma\gamma$ amplitude shows a divergent behavior around
two-particle normal thresholds, related to the presence of square-root and 
logarithmic singularities in the variable $\beta_i^2\equiv 1- 4 M_i^2\slash 
M_H^2$, where $M_H$ is the Higgs mass and $M_i$, with $i=W,Z,t$, stands for
the mass of the $W$ ($Z$) boson or the top quark.

Square-root singularities are related to: 1) derivatives of two-point 
one-loop functions, associated with Higgs wave-function renormalization; 
2) derivatives of three-point one-loop integrals, generated by mass 
renormalization; 3) genuine irreducible two-loop diagrams containing a 
one-loop self-energy insertion.

Concerning the Higgs wave-function renormalization factor at one loop,
it is possible to explicitly prove that the coefficient of the derivative
of the two-point one-loop integral involving a top quark contains a positive 
power of the threshold factor $\beta_t$. Therefore, square-root singularities 
associated with wave-function renormalization appear only at the $2 M_W $ 
and $2 M_Z$ thresholds.

\begin{wrapfigure}{r}{0.6\columnwidth}
     \centerline{\includegraphics[width=0.6\columnwidth]{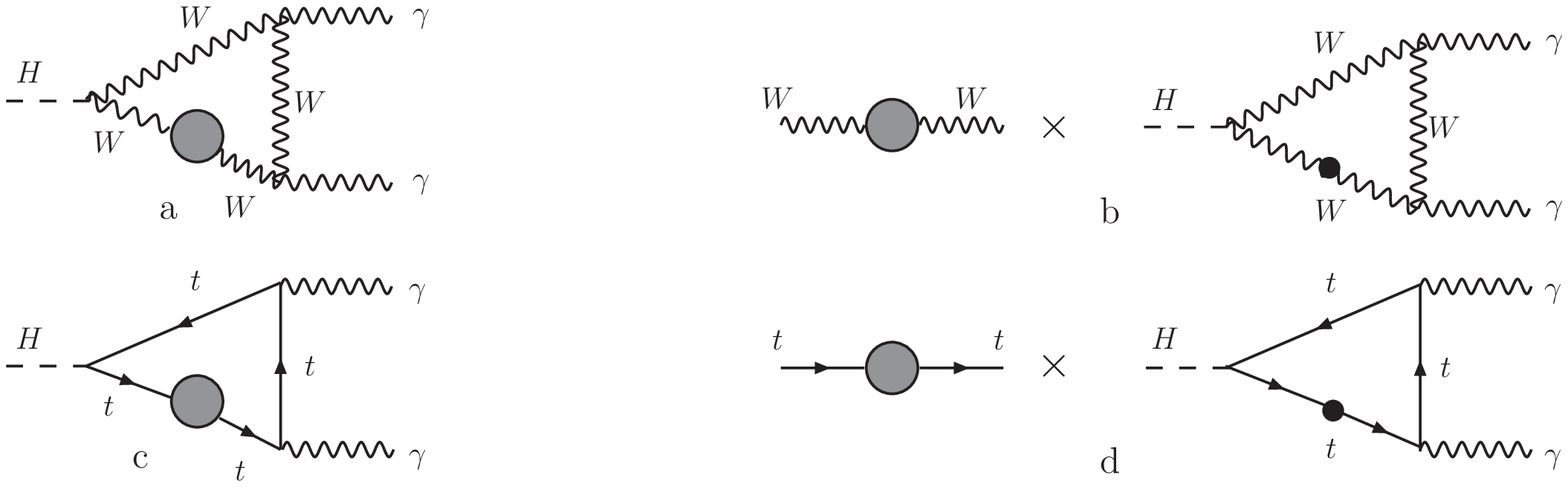}}
     \caption{ two-loop ({\rm a,c}) and mass-renormalization ({\rm b,d})
               diagrams relevant for the analysis of square-root divergencies. 
               Gray circles stand for self-energy insertions, black dots 
               denote derivatives. }
     \label{fig1}
\end{wrapfigure}

We consider now genuine two-loop diagrams containing a self-energy insertion;
they naturally join terms induced by one-loop mass renormalization as shown
in Figure~\ref{fig1}, where bosonic and fermionic diagrams are illustrated.
Fermionic diagrams of Figure~\ref{fig1}c and Figure~\ref{fig1}d are 
$\beta_t$-protected at threshold, and do not require any special care. The 
two-loop irreducible diagram of Figure~\ref{fig1}a can be cast in a representation 
where the divergent part is completely written in terms of the one-loop diagrams 
of Figure~\ref{fig1}b. Moreover, it is possible to check explicitly that the
$\beta_W^{-1}$ behavior  generated by the two-loop diagram exactly cancels the 
$\beta_W^{-1}$ divergency due to one-loop $W$-mass renormalization performed in 
the complex-mass scheme.

Therefore, at the amplitude level, square-root singularities are confined
to the Higgs-boson wave-function renormalization factor (see 
also ~\cite{Kniehl:2001ch}).

\begin{wrapfigure}{l}{0.7\columnwidth}
     \centerline{\includegraphics[width=0.63\columnwidth]{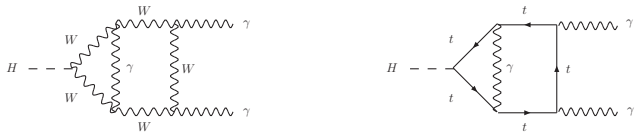}}
     \caption{ irreducible two-loop diagrams which potentially lead to 
               a logarithmic divergency. }
     \label{fig2}
\end{wrapfigure}

As thoroughly discussed in~\cite{Passarino:2007fp,Actis:2008ts}, logarithmic
singularities are generated by the bosonic diagram of Figure~\ref{fig2},
whereas top-quark terms are $\beta_t$-protected at threshold.

\begin{wrapfigure}{r}{0.43\columnwidth}
     \centerline{\includegraphics[width=0.46\columnwidth]{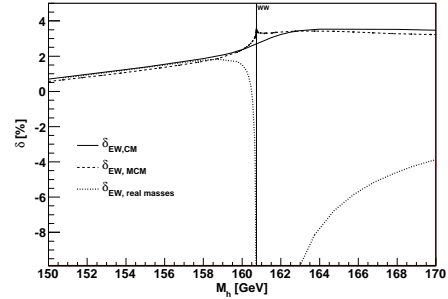}}
     \caption{ percentage NLO electroweak corrections to $\Gamma(H\to\gamma
               \gamma)$ in the MCM and CM setups described in the text. The result 
               obtained using real masses ($\delta_{\rm EW,\, real\, masses}$)
               as inputs is also shown. }
     \label{fig3}
\end{wrapfigure}

A pragmatic gauge-invariant solution to the problem of threshold singularities due 
to unstable particles for the $H\to \gamma \gamma$ decay has been introduced and 
formalized in~\cite{Passarino:2007fp}. In this setup (Minimal Complex-Mass scheme) 
the amplitude is written as the sum of divergent and finite terms, and the 
complex-mass scheme of~\cite{Denner:2005es} is introduced for all gauge-invariant
divergent terms. The Minimal Complex-Mass (MCM) scheme allows for a straightforward 
removal of unphysical infinities. Real masses of unstable gauge bosons are 
traded for complex poles in divergent terms, also at the level of the couplings, 
gauge-parameter invariance and Ward identities are preserved and the amplitude 
has a decent threshold behavior, as shown in Figure~\ref{fig3} for the NLO electroweak 
corrections to the $H\to \gamma \gamma$ decay width ($\delta_{\rm EW,\, MCM}$).

However, the MCM scheme does not deal with artificial cusps associated
with the crossing of gauge-boson normal thresholds, as shown in Figure~\ref{fig3}
for the $WW$ threshold. In order to cure these effects, in~\cite{Actis:2008ts}
we have performed a full implementation of the complex-mass
scheme in the two-loop computation. In the complete Complex-Mass (CM) setup, 
real masses are replaced with complex poles and the real part of the $W$-boson 
self-energy, stemming from mass renormalization at one loop, is traded for the 
full expression, including its imaginary part. As shown in Figure~\ref{fig3},
the full introduction of the CM scheme leads to a complete smoothing of the 
corrections ($\delta_{\rm EW,\, CM}$) at threshold. A more striking effect
is associated with the production of a Higgs boson through gluon-gluon
fusion as discussed in~\cite{Actis:2008ts}.

\section{Results}
\label{sec:num}

\begin{figure}[h]
     \centerline{\includegraphics[width=0.6\columnwidth]{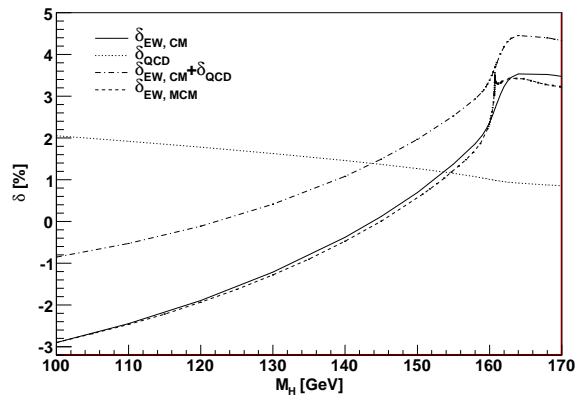}}
     \caption{ NLO percentage corrections to the
               $H\to\gamma\gamma$ decay; see text for details. }
     \label{fig4}
\end{figure}

Numerical results for the percentage NLO electroweak corrections to the partial 
width $\Gamma(H\to\gamma\gamma)$ are shown in Figure~\ref{fig4}. All light-fermion 
masses have been set to zero in the collinear-free amplitude and we have defined 
the $W$- and $Z$-boson complex poles by
\begin{equation*}
  s_{j} \equiv \mu_{j}\,\left(\mu_{j} - i\,\gamma_{j}\right), \quad 
  \mu^2_{j} \equiv M^2_j - \Gamma^2_{j},                      \quad
  \gamma_{j} \equiv \Gamma_{j}\,\left[ 1 - \Gamma^2_{j}/(2\,M^2_j)\right],
\end{equation*}
with $j=W,Z$. As input parameters for the numerical evaluation we have used 
the following values for the masses and the widths of the gauge bosons:
$M_W = 80.398\,{\rm GeV}$, $M_Z = 91.1876\,{\rm GeV}$, $\Gamma_W = 2.093\,{\rm GeV}$ 
and $\Gamma_Z = 2.4952\,{\rm GeV}$.

In Figure~\ref{fig4} we have shown  QCD corrections ($\delta_{\rm QCD}$), electroweak
contributions in the MCM and CM setups ($\delta_{\rm EW, MCM}$ and $\delta_{\rm EW, CM}$)
and the full NLO prediction involving both electroweak effects
in the complex-mass scheme and QCD ones ($\delta_{\rm EW, CM}$+$\delta_{\rm QCD}$).
We observe that QCD and electroweak 
corrections almost compensate below the $WW$
threshold, as shown by~\cite{Aglietti:2004nj,Degrassi:2005mc}, leading to an overall 
very small correction, well below the expected
$3 \%$ accuracy at the planned linear collider operating in the $\gamma \gamma$
option. Above the $WW$ threshold, instead, both corrections are positive and lead
to a global $4 \%$ effect for $M_H=170$ GeV.

\section{Acknowledgments}

G. Passarino, C. Sturm and S. Uccirati are gratefully acknowledged for the
collaboration on the subject. Diagrams have been drawn with 
         {\sc Axodraw}/{\sc Jaxodraw}~\cite{Vermaseren:1994je}.


\begin{footnotesize}

\end{footnotesize}


\end{document}